\documentclass[prl]{revtex4}
\begin{document}

\title{Long range order in gravity\\}

\author{Giorgio Papini}
\altaffiliation[Electronic address:]{papini@uregina.ca}
\affiliation{Department of Physics and Prairie Particle Physics Institute, University of Regina, Regina, Sask. S4S 0A2, Canada}%

\begin{abstract}
Gravity induced condensation takes the form of momentum alignment in an
ensemble of identical particles. Use is made of a one-dimensional Ising model to calculate the alignment per particle
and the correlation length as a function of the temperature. These  parameters
indicate that momentum alignment is possible in the proximity of some astrophysical objects
and in earth, or near earth laboratories. Momenta oscillations behave as known spin oscillations and
obey identical dispersion relations.

\end{abstract}

\maketitle

\section{1. Introduction}

Important phenomena occur in physics when a continuous symmetry of the energy function is broken. Then an order
parameter appears that causes a collective arrangement of matter, or condensation. This is the case, for instance, of ferromagnetism
and superconductivity where the order parameters are represented respectively by the magnetic moment and the amplitude
of paired electrons \cite{BAX}.

It is not known whether gravity can foster and sustain a collective arrangement of matter. This problem has not so far received
attention, though collective phenomena due to other interactions such as electromagnetism,
in the presence of gravity have been studied \cite{DW,GP,BAK,CAS}.
Gravitational corrections to superconducting parameters like penetration depth, or magnetic flux quantization can be
calculated, but the protagonist in this scenario is still the electromagnetic field.

The more fundamental aspect of the problem, when no agents are present other than gravity and the particles on which gravity
acts, is tackled in what follows. The aim is to investigate whether, on the way to quantum gravity, collective phenomena occur.

Use is made of the external field approximation (EFA)\cite{PAP1,PAP0} that treats gravity as a classical
theory when it interacts with quantum particles. Hopefully, at this stage, one
need not necessarily pass through quantum field theory.

ESA can be applied successfully to all those problems
involving gravitational fields of weak to intermediate strength for which the full-fledged use of general relativity
is not required \cite{PASP,PAP5, LAMB,PAP6,PAP8, PAP9,PAP7}. EFA is encountered in the solution of relativistic wave equations and takes slightly different forms
according to the statistics obeyed by the particles \cite{PAP2,PAP3,PAP4}. It can also be applied
to theories in which acceleration has an upper limit \cite{CAI1,CAI2,CAI3,BRA,MASH1,MASH2,MASH3,TOLL,SCHW,PU} and that allow for the resolution
of astrophysical and cosmological singularities in quantum gravity \cite{ROV,BRU} and to those theories of asymptotically
safe gravity that can be expressed as Einstein gravity coupled to a scalar field \cite{CA}.
In particular, EFA can produce results complementary to those of the method
of space-time deformation \cite{CapLamb}.

For the sake of completeness, some essential points are being repeated.

Consider scalar particles first. It has already been shown \cite{PAP10,PAP11} that the covariant
Klein-Gordon (KG) equation gives rise to classical objects that have a vortical structure. This result can be rapidly derived.
Neglecting curvature dependent terms and applying the Lanczos-De Donder condition
\begin{equation}\label{L}
  \gamma_{\alpha\nu,}^{\,\,\,\,\,\,\,
  \nu}-\frac{1}{2}\gamma_{\sigma,\alpha}^\sigma = 0 \,,
  \end{equation}
the covariant KG equation can be re-written, to $\mathcal{O}(\gamma_{\mu\nu})$, in the form
\begin{equation}\label{KG}
\left(\nabla_{\mu}\nabla^{\mu}+m^2\right)\phi(x)\simeq\left[\eta_{\mu\nu}\partial^{\mu}\partial^{\nu}+m^2
+\gamma_{\mu\nu}\partial^{\mu}\partial^{\nu}
\right]\phi(x)=0\,.
\end{equation}
The metric deviation is $\gamma_{\mu\nu}=g_{\mu\nu}-\eta_{\mu\nu}$ and the Minkowski metric $\eta_{\mu\nu}$ has signature $-2$.
Units $\hbar=c=k_{B}=1$ are used unless specified otherwise. The notations are as in \cite{PAP5}.
In particular, partial derivatives with respect to a variable $y_{\mu}$ are interchangeably
indicated by $\partial_{\mu}$, or by a comma followed by $\mu$.
The first order solution of (\ref{KG}) is
\begin{equation}\label{PHA}
\phi(x)=\left(1-i\hat{\Phi}_{G}(x)\right)\phi_{0}(x)\,,
\end{equation}
where $\hat{\Phi}_{G}$ is the operator
\begin{equation}\label{PHI}
\hat{\Phi}_{G}(x)=-\frac{1}{2}\int_P^x
dz^{\lambda}\left(\gamma_{\alpha\lambda,\beta}(z)-\gamma_{\beta\lambda,\alpha}(z)\right)
\left(x^{\alpha}-z^{\alpha}\right)\hat{k}^{\beta}
\end{equation}
\[+\frac{1}{2}\int_P^x dz^{\lambda}\gamma_{\alpha\lambda}(z)\hat{k}^{\alpha}=\int_{P}^{x}dz^{\lambda}\hat{K}_{\lambda}(z,x)\,,\]
where $P$ is an arbitrary point, henceforth dropped, and
$\phi_{0}(x)$ is a wave packet solution of the free KG equation
\begin{equation}\label{KG0}
\left(\partial_{\mu}\partial^{\mu}+m^2\right)\phi_{0}(x)=0 \,.
\end{equation}
The transformation (\ref{PHA}) that makes the ground state of the system space-time dependent,
produces a breakdown of the vacuum symmetry.

For simplicity, a plane wave is chosen for $\phi_{0}$ and one can also write $\hat{\Phi}_{G}(x)\phi_{0}(x)\equiv \Phi_{G}(x)\phi_{0}(x)$,
$\hat{k}_{\alpha}\phi_{0}=i\partial^{\alpha}\phi_{0}=k^{\alpha}\phi_{0}$, where the wave vector $k_{\alpha}$ satisfies the condition $k_{\alpha}k^{\alpha}=m^2$.
$\Phi_{G}$ is the object of primary importance in this work.
The two-point vector $K_{\lambda}(z,x)$, defined by
\begin{equation}\label{K}
K_{\lambda}(z,x)=-\frac{1}{2}\left[\left(\gamma_{\alpha\lambda,\beta}(z)
-\gamma_{\beta\lambda,\alpha}(z)\right)\left(x^{\alpha}-z^{\alpha}\right)-\gamma_{\beta\lambda}(z)\right]k^{\beta}\,,
\end{equation}
is linked to $\Phi_{G}$ by the relation $\Phi_{G}(x)=\int^{x}dz^{\lambda}K_{\lambda}(z,x)$
and is obviously classical. By differentiating (\ref{K}) with
respect to $z^{\alpha}$, one finds \cite{PAP11}
\begin{equation}\label{FT}
\tilde{F}_{\mu\lambda}(z,x)\equiv K_{\lambda,\mu}(z,x)-K_{\mu,\lambda}(z,x)=R_{\mu\lambda\alpha\beta}(z) J^{\alpha\beta}\,,
\end{equation}
where
$R_{\alpha\beta\lambda\mu}(z)=-\frac{1}{2}\left(\gamma_{\alpha\lambda,\beta\mu}
+\gamma_{\beta\mu,\alpha\lambda}-\gamma_{\alpha\mu,\beta\lambda}-\gamma_{\beta\lambda,\alpha\mu}\right)$
is the linearized Riemann tensor satisfying the identity
$R_{\mu\nu\sigma\tau}+R_{\nu\sigma\mu\tau}+R_{\sigma\mu\nu\tau}=0$
and
$J^{\alpha\beta}=\frac{1}{2}\left[\left(x^{\alpha}-z^{\alpha}\right)k^\beta-k^\alpha
\left(x^\beta-z^\beta\right)\right]$ is the angular momentum about
the base point $x^\alpha$. The Maxwell-type equations
\begin{equation}\label{ME1}
\tilde{F}_{\mu\lambda,\sigma}+\tilde{F}_{\lambda\sigma,\mu}+\tilde{F}_{\sigma\mu,\lambda}=0
\end{equation}
and
\begin{equation}\label{ME2}
\tilde{F}^{\mu\lambda}_{\,\,\,\,\,\,\,,\lambda}\equiv -j^{\mu}=
\left(R^{\mu\lambda}_{\,\,\,\,\,\,\,\alpha\beta}J^{\alpha\beta}\right),_{\lambda}
=R^{\mu\lambda}_{\,\,\,\,\,\,\,\alpha\beta,\lambda}\left(x^\alpha
-z^\alpha\right)k^\beta +R^{\mu}_{\,\,\,\,\beta}k^{\beta}\,,
\end{equation}
can be obtained from (\ref{FT}) using the Bianchi identities $R_{\mu\nu\sigma\tau,\rho}
+R_{\mu\nu\tau\rho,\sigma}+R_{\mu\nu\rho\sigma,\tau}=0$. The current $j^{\mu}$ satisfies the conservation law
$j^{\mu}_{\,\,\,,\mu}=0$.
Equations (\ref{ME1}) and (\ref{ME2}) are identities and do not represent additional constraints on $\gamma_{\mu\nu}$.
They hold true, in EFA, for any metrical field theory.
The vector $K_\lambda$ is non-vanishing only on surfaces $\tilde{F}_{\mu\nu}$ that
satisfy (\ref{ME1}) and (\ref{ME2})
and represent the vortical structures generated by $\Phi_{G}$.
At a point $z_{\alpha}$ along the path
\begin{equation}\label{PD}
\frac{\partial \Phi_{G}(z)}{\partial z^{\sigma}}=-\frac{1}{2}\left[\left(\gamma_{\alpha\sigma,\beta}(z)-\gamma_{\beta\sigma,\alpha}(z)\right)
\left(x^{\alpha}-z^{\alpha}\right)-\gamma_{\beta\sigma}(z)\right]k^{\beta}=K_{\sigma}(z)\,,
\end{equation}
and
\begin{equation}\label{PD2}
\frac{\partial^{2}\Phi_{G}(z)}{\partial z^{\tau}\partial z^{\sigma}}-\frac{\partial^{2}\Phi_{G}(z)}{\partial z^{\sigma}\partial z^{\tau}}=R_{\alpha\beta\sigma\tau}\left(x^{\alpha}-z^{\alpha}\right)k^{\beta}\equiv\left[\partial z_{\tau},\partial z_{\sigma}\right]\Phi_{G}(z)=
\tilde{F}_{\tau\sigma}(z)\,.
\end{equation}
It follows from (\ref{PD2}) that $\Phi_{G}$ is not single-valued and that, after a gauge transformation, $K_{\alpha}$  satisfies the equations
\begin{equation}\label{DIV}
\partial_{\alpha}K^{\alpha}=\frac{\partial^{2}\Phi_{G}}{\partial z_{\sigma}\partial z^{\sigma}}=0
\end{equation}
and
\begin{equation}\label{2der}
\partial^2 K_{\lambda}=-\frac{k^{\beta}}{2}\left[\left(\partial^2 (\gamma_{\alpha\lambda,\beta})-\partial^2(\gamma_{\beta\lambda,\alpha})\right)\left(x^\alpha -z^\alpha\right)+
\partial^2 \gamma_{\beta\lambda}\right]
\end{equation}
identically, while the equation
\begin{equation}\label{PD3}
\left[\partial z_{\mu},\partial z_{\nu}\right]\partial z_{\alpha}\Phi_{G}=-\left(\tilde{F}_{\mu\nu,\alpha}+\tilde{F}_{\alpha\mu,\nu}+\tilde{F}_{\mu\alpha,\nu}\right)=0\,,
\end{equation}
holds everywhere. Therefore, the potential $K_{\alpha}$ is regular everywhere, which is physically desirable, but $\Phi_{G}$ is
singular. It also follows from (\ref{PD2}) that $\tilde{F}_{\mu\nu}$ is a vortex along which
the scalar particles are dragged with acceleration
\begin{equation}\label{GE}
\frac{d^{2}z_{\mu}}{ds^{2}}=u^{\nu}\left(u_{\mu,\nu}-u_{\nu,\mu}-R_{\mu\nu\alpha\beta}\left(x^{\alpha}-z^{\alpha}\right)u^{\beta}\right)\,,
\end{equation}
and relative acceleration
\begin{equation}\label{GD}
\frac{d^{2}(x_{\mu}-z_{\mu})}{ds^{2}}=\tilde{F}_{\mu\lambda}u^{\lambda}=R_{\mu\beta\lambda\alpha}\left(x^{\alpha}-z^{\alpha}\right)u^{\beta}u^{\lambda}\,,
\end{equation}
in agreement with the equation of geodesic deviation \cite{PAP10,PAP11}.

From (\ref{PHI}) and $\phi_{0}$ one can derive the important relation
\begin{equation}\label{PHMU}
k^{\mu}\Phi_{G,\mu}=1/2 \gamma_{\mu\nu}k^{\mu}k^{\nu}\,,
\end{equation}
by straight differentiation of $\Phi_{G}$. It is also useful to recall that the momentum of $\phi(x)$ is \cite{PAP5}
\begin{equation}\label{MM}
P_{\mu}=k_{\mu}+\Phi_{G,\mu}\,,
\end{equation}
and that the solution (\ref{PHA}) also preserves its structure at higher order iterations
according to the relation
\begin{equation}\label{PHIn}
\phi(x)=\Sigma_{n}\phi_{(n)}(x)=\Sigma_{n}e^{-i\hat{\Phi}_{G}}\phi_{(n-1)}\,.
\end{equation}

The plan of the paper is the following. It is shown in Section 2 that (\ref{PHA}) does indeed lead to the formation
of Nambu-Goldstone bosons as it should when a symmetry is broken. The order parameter is identified with $\Phi_{G,\mu}$
and its excitations are further studied in Section 3 by means of a lattice gas model. A correlation length is then calculated.
It shows that the fluctuations of the order parameter are correlated over distances higher than the lattice spacing.
The oscillations of $P_{\mu}$, or, equivalently of $k_{\alpha}$ are studied in Section 4. The
results are summarized and discussed in Section 5.

\section{2. Nambu-Goldstone bosons}

Nambu-Goldstone bosons are particles that arise when a continuous symmetry is broken, as in the case of (\ref{PHA}).
By substituting (\ref{PHA}) into the energy function $\tilde{H}=g_{\mu\nu}(k^{\mu}+\nabla^{\mu}\Phi_{G}(x))(k^{\nu}+\nabla^{\nu}\Phi_{G})$ and
using (\ref{PHMU}), one gets, to $\mathcal{O}(\gamma_{\mu\nu})$,
\begin{equation}\label{EXC}
\tilde{H}=g_{\mu\nu}P^{\mu}P^{\nu}=m^{2}+2\gamma_{\mu\nu}k^{\mu}k^{\nu}\,,
\end{equation}
that is, the fluctuations $\Phi_{G}$ about the original symmetric state $\phi_{0}$ are equivalent to the production of
a Nambu-Goldstone boson that is massless because $\partial^{2}\Phi_{G}/(\partial z_{\sigma}\partial z^{\sigma})=\partial_{\alpha}K^{\alpha}=0$.
The field $K_{\lambda}$ satisfies the equation \cite{PAP11}
\begin{equation}\label{2der}
\partial^2 K_{\lambda}=-\frac{k^{\beta}}{2}\left[\left(\partial^2 (\gamma_{\alpha\lambda,\beta})-\partial^2(\gamma_{\beta\lambda,\alpha})\right)\left(x^\alpha -z^\alpha\right)+
\partial^2 \gamma_{\beta\lambda}\right]
\end{equation}
identically. If $ \partial^{2}\gamma_{\alpha\lambda,\beta}\neq 0 $, then $K_{\lambda}$ acquires a mass.
If $\partial^{2}\gamma_{\alpha\beta}=0$, then $\partial^{2}K_{\lambda}=0$ and $K_{\lambda}(x)$ remains massless.
Equation (\ref{2der}) therefore ensures that no physical degrees of freedom are gained or lost in the
rearrangement of symmetry leading from $\gamma_{\mu\nu}$ to $K_{\alpha}$.

Though use is not made  of this possibility in the present work, the addition to $j_{\alpha}$ of a current proportional
to $K_{\alpha}$ does not violate
the equation of conservation $j^{\alpha}_{,\alpha}=0$ because $K^{\alpha}_{\,\,,\alpha}=0$. In this case
gravitons acquire a mass.

Expressions similar to (\ref{MM}) can be obtained for the generalized momenta of spin-1 \cite{PAP3} and spin-2 particles \cite{PAP4}, while for fermions one finds \cite{PAP5}
\begin{equation}\label{FF}
P^{f}_{\mu}=k_{\mu}+\Phi_{G,\mu}+\Gamma_{\mu}\,,
\end{equation}
where the additional term $\Gamma_{\mu}(x)=\frac{i}{4}\gamma^{\nu}(x)(\nabla_{\mu}\gamma_{\nu}(x))$ represents the spinorial connection.
The spin contributions to $P_{\mu}$ are neglected wherever the spin of the particle is much less than its angular momentum.

In condensed matter physics, the low energy excited states can take, for instance, the appearance of phonons
in crystals when translational invariance is broken while spin waves (magnons) result from a spontaneously broken rotational symmetry of a
ferromagnet. Order parameters appear: for ferromagnets the order parameter is the average magnetic moment.
There are condensation forces associated with the order parameter like the magnetic field for magnons \cite{BAX}.
In the present problem one can derive from (\ref{PHMU}) and (\ref{MM}) the relation
\begin{equation}\label{M}
\frac{1}{2}\gamma_{\mu\nu}k^{\mu}k^{\nu}=\left(P_{\nu}-k_{\nu}\right)k^{\nu}=k^{\mu}\Phi_{G,\mu}=\frac{d\Phi_{G}}{ds}\,,
\end{equation}
where $s$ is the affine parameter along the world-line of $m$. It therefore seems natural to identify $\Phi_{G,\mu}$ with
the order parameter and $\frac{d\Phi_{G}}{ds}$ with the condensation force.
It follows that the phenomenon of condensation corresponds
to the alignment of the momenta $k_{\mu}$ with $P_{\mu}$ which represents the lowest energy configuration of the system.
$\Phi_{G,\mu}$ is therefore responsible for momentum alignment and for the reduction of $\gamma_{\mu\nu}$ to $K_{\alpha}$.
It can be verified that the equivalence principle still holds true along the worldline of $m$. In fact
\begin{equation}\label{X}
\frac{d^{2}x_{\mu}}{ds^{2}}=\Phi_{G,\mu\nu}u^{\nu}=\left(\frac{d\Phi_{G}}{ds}\right)_{,\mu}=\Gamma^{\alpha}_{\mu\nu}u_{\alpha}u^{\nu}\,,
\end{equation}
which does not depend on $m$.

The energy function is usually given in units of energy rather than energy square. Dropping the
unnecessary term $m$, equation (\ref{EXC}) gives
\begin{equation}\label{HE}
H= -\frac{1}{m}\gamma_{\mu\nu}k^{\mu}k^{\nu}\,,
\end{equation}
for low energy particles. The lowest energy state is the one in which all momenta point in the same direction.

For particles of small mass, (\ref{HE}) can be replaced by $H=-(1/E)\gamma_{\mu\nu}k^{\mu}k^{\nu}$,
where $E\sim k_{0}$ is the free energy of the particle.

\section{3. The lattice gas model}

The properties of a system of a large number of particles satisfying (\ref{PHA}) follow from
$H$ which strongly resembles the energy function of the Ising model. A difference is represented here by the vectors $k_{\alpha}$
(or $P_{\alpha}$) that replace in (\ref{EXC}) the Ising spin variables
$\sigma_{i}$ which are numbers that can take the values $\pm 1$. It is however known that a
lattice gas model \cite{BAX}, equivalent to the Ising model, can be set up in which
the particles are restricted to lie only on the $N$ sites of a fine lattice, instead of being allowed to occupy any position in space-time.
Then one can associate to each site $i$ a variable $s_{i}=(1+\sigma_{i})/2$ which takes the value $1$ if the site is occupied by a vector $k_{\alpha}$ and
the value $0$ if it is empty. Any distribution of the particles can be indicated by the set of their site occupation numbers
${s_{1},. ...s_{N}}$. The number of particle in this arrangement is $n=s_{1}+s_{2}+...s_{N}$. One can further consider a chain of length N as in
the one-dimensional Ising model. By replacing
$k^{\mu}k^{\nu}$ in (\ref{EXC}) with their average $k^2 \eta^{\mu\nu}/4$ over the angle and restricting the interaction to couples of nearest neighbour sites,
one obtains
\begin{equation}\label{HH}
H=-\frac{m}{4} \gamma\left(\sum_{k=1}^{N}s_{k}s_{k+1}\right)\,,
\end{equation}
where $\gamma\equiv \gamma_{\mu\nu}\eta^{\mu\nu}$. The average over vector directions is obviously unnecessary for conformally flat gravitational fields.
In the case of particles of vanishingly small mass, by averaging over the directions of the Euclidean vectors $k'_{\alpha}$,
one gets $k^{'2}=-k_{0}^{'2}-\vec{k}^{2}\sim -k_{0}^{'2}=E^{2}$
and $E$ replaces $m$ in (\ref{HH}).
By imposing periodic conditions $s_{N+1}=s_{1}$ along the hypercylinder with axis parallel to the time-axis,
the partition function becomes
\begin{equation}\label{Z}
Z=\sum_{s_{1}}...\sum_{s_{N}}\exp\left(\beta \gamma \frac{m}{4}\sum_{k=1}^{N}s_{k}s_{k+1}\right)\,.
\end{equation}
This one-dimensional Ising model has no time-dependent dynamics because it does not specify
how each variable $s_{i}$ varies in time. It is however assumed that $s_{i}$ can be changed at any time
so that statistical mechanics can be applied.
The model can be solved exactly \cite{BAX} by introducing the transfer matrix $<s|\tilde{M}|s'>=exp(\beta \varepsilon ss')$,
where $\beta\equiv1/T$ and $\varepsilon\equiv m \gamma/4$ contains the gravitational contribution due to $\gamma_{\mu\nu}$.
Equation (\ref{Z}) can be rewritten as $Z=\sum_{s_{1}}<s_{1}|\tilde{M}^{N}|s_{1}>=Tr(\tilde{M}^{N})=\lambda_{+}^{N}+\lambda_{-}^{N}$ and
the eigenvalues of $\tilde{M}$ are $\lambda_{+}=2 \cosh(\beta\varepsilon)$ and $\lambda_{-}=2\sinh(\beta\varepsilon)$. As $N\rightarrow \infty$,
only $\lambda_{+}$ is relevant, $N^{-1}\ln Z\rightarrow \ln(\lambda_{+})$ and the Helmholtz free energy per site is
$F/N=-(N \beta)^{-1}\ln Z\rightarrow -\beta^{-1}\ln(\lambda_{+})$. The alignment per particle for large values of $N$ is
\begin{equation}\label{AA}
\Gamma \equiv-\frac{1}{N}\frac{\partial F}{\partial\varepsilon}\sim -\frac{1}{\beta}\frac{d\ln \lambda_{+}}{d\varepsilon}=1-\frac{2e^{-2\beta\varepsilon}}{1+e^{-2\beta\varepsilon}}\,,
\end{equation}
which yields the gravitational correction due to $\varepsilon$. It also follows from (\ref{AA}) that there is no spontaneous momentum alignment ($\Gamma=0$ when $\varepsilon=0$)
and that complete alignment $\Gamma=1$ is possible only for $T\rightarrow 0$.
In fact $F\rightarrow -N\varepsilon $ in the limit $T\rightarrow 0$ for completely aligned momenta and one can say that there is a phase transition
at $T=0$, but none for $T>0$.
It follows from (\ref{AA}) that the value of $\Gamma$ depends on $\gamma$.
It also follows that there is no alignment
($\Gamma =0$) for $T\rightarrow\infty$ (for any $\gamma$ and $m$), or for $\gamma =0$ (no gravity and any $T$).
According to (\ref{AA}), complete alignment $\Gamma =1$ can be achieved only at $T=0$
which plays the role of a critical temperature in the model.

The results given above also apply to a two-dimensional Ising model because the partition function (\ref{Z}) remains unchanged under the same conditions.

Following textbook procedures, one can also calculate the correlation length defined as
\begin{equation}\label{CC}
<s_{0}s_{j}>=\frac{1}{Z}\sum_{s_{j}}s_{0}s_{j}\exp\left\{\beta \sum_{i=1}^{N-1}\varepsilon_{i} s_{i}s_{i+1}\right\}=\frac{1}{Z}\frac{\partial^{j}(\varepsilon_{0}...\varepsilon_{N})}{\partial\varepsilon_{0}...\partial \varepsilon_{j-1}}\,,
\end{equation}
where $s_{0}$ is any particular site chosen far away from the extremes of the chain and the final result is evaluated at $\varepsilon_{i}=\varepsilon$.
The result is $<s_{0}s_{j}>=\left[\tanh\beta\varepsilon\right]^{j} \equiv \exp(-\frac{j}{\xi})$,  where
$\xi=-[\ln(\tanh(\beta \varepsilon)
]^{-1}\approx -2\exp(-2\beta \varepsilon)$ because $\beta\varepsilon<1$ and $\xi>0$.
Finally the correlation length in units of lattice spacing is \cite{BAX}
\begin{equation}\label{CSI}
\xi\sim\frac{1}{2}\exp(2\beta\varepsilon)\,,
\end{equation}
which gives $\xi=\infty$ at $T=0$ and $\xi=0$ at $T=\infty$ where thermal agitation can effectively uncouple neighbouring sites.
The parameter $\xi$ can be large for some astrophysical objects, as shown below.

Some order of magnitude estimates are instructive. For electrons in the neighborhood of a white dwarf
$2\beta\epsilon \sim 2\cdot10^{6}/T$, where $T$ is in Kelvin, while near earth $2 \beta\varepsilon \sim 20/T$. Similarly,
for a nucleon close to a neutron star $2 \beta\varepsilon \sim 6\cdot10^{11}/T$.
Given that for an old white dwarf $T\sim 10^{4}K$ and that for a newly formed neutron star $T\sim 10^{12}$, one finds that $\xi$ can be very large
and that $\Gamma \sim 1$. For electrons at earth one finds $\Gamma\sim 4.8\cdot 10^{-2}$ and $\xi\sim 0.55$. These values increase
at lower temperatures. At liquid He temperatures one obtains $\Gamma\sim 0.99$ and $\xi\sim 1.3\cdot 10^{6}$.

It is also interesting to estimate the velocity reached by a particle of initial velocity $V$ along the circular path.
By assuming for simplicity that $k_{3}=0$, one gets from $\vec{V}=\vec{P}/P_{0}$, $\vec{v}=\vec{k}/k_{0}$ and $P_{i}=k_{i}+\gamma_{i\nu}k^{\nu}$ and
\begin{equation}\label{VV}
V^{2}=v_{1}^{2}+v_{2}^{2}\simeq v_{1}^{2}+v_{2}^{2}+2v_{1}^{2}\gamma_{11}+2v_{2}^{2}\gamma_{22}\simeq v^{2}+2v^{2}\gamma_{11}\,,
\end{equation}
Taking $\gamma_{11}\sim \gamma_{22}$, one finds
\begin{equation}\label{V}
V-v \sim v \gamma_{11}\,.
\end{equation}
Notice that if $v=Ar$, as for a rotating solid body and $\gamma_{11}=GM/r$, one gets $V-v \sim 2AGM$, where $A$
is a constant, which is reminiscent of
the characteristic rotation speed of galaxies, at least in a certain range of distances from the nucleus.

\section{4. Momentum oscillations}

The oscillations of $P_{\mu}$ are waves akin to spin waves for magnons for which Kittel has given a
classical derivation \cite{KIT}. The corresponding derivation for the gravitational case is as follows.
The terms that contain a generic $q$ term in (\ref{HH}) are $-(m\gamma/4) \vec{s}_{q}\cdot (\vec{s}_{q-1}+\vec{s}_{q+1})$
which, on account of the equation of deviation (\ref{GD}) can be written as $-(m\gamma/4)\vec{s}_{q}\cdot \vec{H}_{q}$,
where $H_{i}=\epsilon_{ijk}R^{kj}_{\,\,\,\,\alpha\beta}J^{\alpha\beta}$
represents the "magnetic" components of $F_{\mu\nu}$. Using simple algebra it is possible to show that
$\vec{s}_{q}\times \vec{H}_{q}=-\vec{s}_{q}\cdot \vec{H}_{q}$ which must then equal the change in momentum
\begin{equation}\label{OSC}
\frac{d\vec{s}_{q}}{dt}=\frac{m\gamma}{4}\vec{s}_{q}\times \left(\vec{s}_{q-1}+\vec{s}_{q+1}\right)\,.
\end{equation}
By assuming that the excitation has small amplitude so that $s_{q}^{x,y}\ll s$, where $s$ is the ground state value
of $s_{k}$, one finds $ds_{q}^{z}/dt=0$,
\begin{equation}\label{sx}
\frac{ds_{q}^{x}}{dt}=\frac{ms\gamma}{4}\left(2s_{q}^{y}-s^{y}_{q-1}-s^{y}_{q+1}\right)
\end{equation}
and
\begin{equation}\label{sy}
\frac{ds_{q}^{y}}{dt}=-\frac{ms\gamma}{4}\left(2s_{q}^{x}-s^{x}_{q-1}-s^{x}_{q+1}\right)\,,
\end{equation}
which have the solutions $s^{x}_{q}=u\cos(qka-\omega t)$ and $s^{y}_{q}=u\sin(qka-\omega t)$,
where $u$ is a constant and $a$ the lattice constant,
provided the determinant of the coefficients of (\ref{sx}) and (\ref{sy}) vanishes. This condition yields the
dispersion relation
\begin{equation}\label{DR}
\omega =\frac{ms\gamma}{2}\left(1-\cos ka\right)\,,
\end{equation}
which coincides with the result of reference \cite{KIT} for magnons.

\section{5. Summary and conclusions}

Condensation appears in physics when a symmetry of the energy function is broken. The fact that gravity breaks the symmetry
of the vacuum is in itself a noticeable result. It does so in EFA for general relativity as well as for any metrical theory,
as a consequence of the term $\gamma_{\mu\nu}k^{\mu}k^{\nu}$ which is classical in origin and appears in all
covariant wave equations. This leads to gravitational effects of a new type like long range order
and momenta alignment that obey temperature laws.
The symmetry changes
introduced  by  solutions of the type (\ref{PHA}) that occur in connection with covariant wave equations, (\ref{KG}) in
particular, have been studied in some detail by making use of EFA that has known applications and iteration procedures and in which
gravity appears through $K_{\alpha}$. This field carries information about matter through $k_{\beta}$, as shown by
(\ref{K}) and, upon quantization would acquire the characteristics of a quasiparticle. Nambu-Goldstone
bosons appear, as indicated by (\ref{EXC}).

The rearrangement of symmetry regarding $\gamma_{\mu\nu}$ satisfies
(\ref{2der}) so that $K_{\alpha}$ has the same degrees of freedom of $\gamma_{\mu\nu}$. According to (\ref{M}), $\Phi_{G}$
is subjected, along a particle worldline, to a "condensation force" that is independent of mass. Symmetry breaking is also accompanied
by a topological change in space-time because $\Phi_{G}$ is multivalued \cite{PAP11}, a phenomenon
discussed at length in \cite{CASE,COH}.

The energy functions are proportional
to $\gamma_{\mu\nu}k^{\mu}k^{\nu}$ and are therefore similar to those used in the Ising model. The consequences
of (\ref{HE}) have then been studied by means of a lattice gas model applied to a one-dimensional chain of lattice sites. This is the simplest
and most direct approach. The model has an exact solution that shows that the alignment per
particle (condensation along $P_{\mu}$) is complete for $T=0$ and vanishing for
$T\rightarrow\infty $ and that alignment can occur only in the presence of gravity ($\gamma_{\alpha}^{\alpha}\neq 0$).
These results are unexpected because, unlike ferromagnetism that deals with the ordering of magnetic moments,
gravitational dipoles do not seem to exist in nature.

Similar results can be obtained for a two-dimensional chain because the partition function $Z$ remains unchanged in this case.
The oscillations of $P_{\mu}$, or equivalently, of the vectors $k_{\mu}$, have behaviour similar to those of magnons and
identical dispersion relations.

The XY-model can be easily recovered from (\ref{HE}) by setting the angles that the vectors $k^{\alpha}$
make with the time-axis and the colatitudes equal to $\pi/2$. Fluctuations have an important relationship
with the dimensionality of space-time and results in higher dimensions
must be confirmed by suitable calculations.

The correlation length $\xi$ shows that correlations of the Ising variables extend
over distances that are large relative to the lattice spacing.
The result is that $\xi=\infty$ at $T=0$ and $\xi=0$ at $T=\infty$ where the thermal agitation becomes extremely large.

Crude order of magnitude estimates have been calculated for white dwarfs and neutron stars. The results yield
reasonable values of $\Gamma $ and $\xi $ and indicate that
gravity induced condensation is possible in the proximity of white dwarfs and neutron stars and for temperatures
characteristic of these astrophysical sources.
Earth values are encouraging and suggest that the effect might be observable in earth, or
near earth laboratories.

The velocity of a particle along the lattice chain has also been calculated. If the initial velocity $v \propto r $,
the typical velocity for  a rotating solid sphere, then $V$ acquires the characteristic velocity of galactic rotation curves
a few kpc from the nucleus.

 Gravity induces condensation in ensembles of identical particles. Particle condensation takes the
form of momentum alignment controlled by temperature dependent parameters like alignment per particle and correlation
length.

 The results may have a bearing on a variety of topics of current interest in astrophysics, cosmology,
space-time deformation and topological structure, the genesis and development of compact objects and galaxies and space-time structure and topology.

\end{document}